\documentclass{aa}

\newcommand{\msun}{\,\mathrm{M}_\odot}

\newcommand{\kms}{\,\mathrm{km}\,\mathrm{s}^{-1}}

\newcommand{\au}{\,\mathrm{AU}}

\newcommand{\yr}{\,\mathrm{yr}}
\newcommand{\kyr}{\,\mathrm{kyr}}

\newcommand{\pc}{\,\mathrm{pc}}

\def\lesssim{\mathrel{\hbox{\rlap{\hbox{\lower3pt\hbox{$\sim$}}}\hbox{\raise2pt\hbox{$<$}}}}}
\def\lesseq{\mathrel{\hbox{\rlap{\hbox{\lower3pt\hbox{$-$}}}\hbox{\raise2pt\hbox{$<$}}}}}
\def\gtrsim{\mathrel{\hbox{\rlap{\hbox{\lower3pt\hbox{$\sim$}}}\hbox{\raise2pt\hbox{$>$}}}}}
\def\gtreq{\mathrel{\hbox{\rlap{\hbox{\lower3pt\hbox{$-$}}}\hbox{\raise2pt\hbox{$>$}}}}}

\usepackage{xcolor}
\usepackage{graphicx}
\usepackage{subcaption}
\usepackage{txfonts}
\usepackage{natbib,twoopt}
\usepackage[breaklinks=true]{hyperref} 
\bibpunct{(}{)}{;}{a}{}{,}
\usepackage{cleveref}
\usepackage{todonotes}
\usepackage{multicol}
\makeatletter
  \newcommandtwoopt{\citeads}[3][][]{\href{http://adsabs.harvard.edu/abs/#3}%
    {\def\hyper@linkstart##1##2{}%
     \let\hyper@linkend\@empty\citealp[#1][#2]{#3}}}
  \newcommandtwoopt{\citepads}[3][][]{\href{http://adsabs.harvard.edu/abs/#3}%
    {\def\hyper@linkstart##1##2{}%
     \let\hyper@linkend\@empty\citep[#1][#2]{#3}}}
  \newcommandtwoopt{\citetads}[3][][]{\href{http://adsabs.harvard.edu/abs/#3}%
    {\def\hyper@linkstart##1##2{}%
     \let\hyper@linkend\@empty\citet[#1][#2]{#3}}}
  \newcommandtwoopt{\citeyearads}[3][][]%
    {\href{http://adsabs.harvard.edu/abs/#3}
    {\def\hyper@linkstart##1##2{}%
     \let\hyper@linkend\@empty\citeyear[#1][#2]{#3}}}
\makeatother
\begin{document}

   \title{The contribution of binary star formation via core fragmentation on protostellar multiplicity}
   \titlerunning{Simulating multiplicity statistics}


   \author{Rajika L. Kuruwita
          \inst{1, 2},
          \and
          Troels Haugb{\o}lle\inst{2}
          }

    \institute{Heidelberg Institute for Theoretical Studies, Schlo{\ss}-Wolfsbrunnenweg 35, 69118 Heidelberg, Germany\\
              \email{rajika.kuruwita@h-its.org}
         \and
             Niels Bohr Institute,
              University of Copenhagen, {\O}ster Voldgade 5-7, DK-1350, Copenhagen K, Denmark\\
             }

   \date{Received September 9, 2022; accepted May 10, 2023}

 
  \abstract
   {Observations of young multiple star systems find a bimodal distribution in companion frequency and separation. The origin of these peaks has often been attributed to binary formation via core and disc fragmentation. However, theory and simulations suggest that young stellar systems that form via core fragmentation undergo significant orbital evolution.}
   {We investigate the influence of the environment on the formation and orbital evolution of multiple star systems, and how core fragmentation contributes to the formation of close ($20-100\au$) binaries. We use multiple simulations of star formation in giant molecular clouds and compare them to the multiplicity statistics of the Perseus star-forming region.}
    {Simulations were run with the adaptive mesh refinement code \texttt{RAMSES} with sufficient resolution to resolve core fragmentation beyond $400\au$ and dynamical evolution down to $16.6\au$, but without the possibility of resolving disc fragmentation. The evolution of the resulting stellar systems was followed over millions of years.}
   {We find that star formation in lower gas density environments is more clustered; however, despite this, the fractions of systems that form via dynamical capture and core fragmentation are broadly consistent at $\sim$40\% and $\sim$60\%, respectively. In all gas density environments, we find that the typical scale at which systems form via core fragmentation is $10^{3-3.5}\au$. After formation, we find that systems that form via core fragmentation have slightly lower inspiral rates ($\sim10^{-1.68}\,\mathrm{AU/yr}$ measured over the first $10000\yr$) compared to dynamical capture ($\sim10^{-1.32}\,\mathrm{AU/yr}$). We then compared the simulation with the conditions most similar to the Perseus star-forming region to determine whether the observed bimodal distribution can be replicated. We find that it can be replicated, but it is sensitive to the evolutionary state of the simulation.}
   {Our results indicate that a significant number of low-mass close binaries
   with separations from $20 - 100\au$ can be produced via core fragmentation or dynamical capture due to efficient inspiral, without the need for a further contribution from disc fragmentation.}


   \maketitle
%

\section{Introduction}

Around half of all stars exist in binary or multiple star systems \citep{moe_mind_2017}, and many are born with a companion \citep{chen_sma_2013, offner_origin_2022}. Models of star and planet formation must account for this multiplicity.

The main proposed pathways for binary star formation are disc fragmentation, core fragmentation, and dynamical capture. Disc fragmentation ($a \lesssim 100\au$) can occur in massive, gravitationally unstable (i.e. Toomre Q parameter $<1$) protostellar discs, if the gravitationally unstable regions can efficiently cool \citep{gammie_nonlinear_2001}. This has been modelled in simulations \citep{takaishi_new_2021}, and with high angular resolution imaging, it is possible to see what could be the outcome of disc fragmentation observationally \citep{tobin_triple_2016}. \citet{bate_statistical_2019} investigated the dependency of metallicity on fragmentation and find in their models that  $15-20\%$ of systems formed via this pathway.
    
Core fragmentation ($100\au \lesssim a \lesssim 10000\au$) occurs in turbulent protostellar cores. Simulations of core fragmentation into binaries have often displayed significant orbital migration from their initial separation \citep{ostriker_dynamical_1999, wurster_collapse_2018, lee_formation_2019, kuruwita_dependence_2020, saiki_twin_2020}, and separations can shrink due to the ejection of tertiary companions \citep{armitage_ejection_1997} down to $<100\au$. Observations of wide binaries with misaligned discs have also been attributed to formation via core fragmentation \citep{lee_formation_2017}.

Dynamical capture occurs when stars that were initially born single and unbound to any other nearby stellar system, later become part of a bound system \citep{parker_binaries_2014}. There is evidence suggesting a third of multiple stars did not form together and instead became bound via dynamical interactions \citep{murillo_siblings_2016}. This could in part be sourced by primordial wide binaries that formed in clusters that disintegrate and reconfigure into new systems \citep{elliott_crucial_2016}.

Observations of the companion frequency distribution of protostars have found a bimodal distribution \citep{tobin_vla_2016, tobin_vlaalma_2022}. For the Perseus molecular cloud, \citet{tobin_vla_2016} find a peak at approximately $75\au$ and $3000\au$. The authors suggest that the peak at $75\au$ is caused by disc fragmentation, while the peak at $3000\au$ is likely caused by fragmentation of protostellar cores. They hypothesise that core fragmentation is a dominant mechanism for systems with separations $>200\au$. However, the authors also hypothesise that the lack of binaries with separations $>1000\au$ may be due to rapid in-spiralling to lower separations.

Surveys of circumstellar disc sizes find the average radius to be $\leq$75~au \citep{cox_protoplanetary_2017, ansdell_alma_2018} for low-mass young stars and the very large and massive discs that could fragment to form stars are uncommon. While there is evidence that disc fragmentation can form close binaries, it is unlikely to be the sole source of binaries with separations of $\sim$10s of AU.

In this work, we investigate the contributions of core fragmentation and dynamical capture to the origin of the observed bimodal separation distribution. We ran multiple simulations of star formation from giant molecular clouds with different initial masses, to investigate how the environment may influence formation pathways for multiple star systems, and the orbital evolution of these systems.

In \Cref{sec:method} we summarise the \texttt{RAMSES} code and the simulation setup. In \Cref{sec:results} we look at the formation and evolution of the multiple star systems that form, and in \Cref{sec:obs_comparison} we include a comparison with observations. In \Cref{sec:discussion} we discuss the overall results of this work and the implications. In \Cref{sec:caveats} we discuss some of the limitations of this paper and suggestions for future work.

\begin{table}
    \caption{Simulation properties}
    \centering
    \begin{tabular}{p{0.6cm}p{1.55cm}p{0.5cm}p{1.5cm}p{1.75cm}}
        \hline
        $M_{gas}$ & $T_{end}$ ($T_{5}$) & SFE & $M_\star$ ($M_{\star,5}$) & $N_\star$ ($N_{\star,5}$)  \\
        M$_\odot$ & Myr & \% & M$_\odot$ & \\
        \hline
        1500 & 3.97 (3.95) & 5 & 75 (75) & 86 (86)\\
        3000 & 2.56 (1.43) & 13 & 393 (150) & 413 (191) \\
        3750 & 1.36 (1.05) & 7 & 252 (187) & 411 (313) \\
        4500 & 1.09 (0.80) & 8 & 341 (225) & 506 (394) \\
        6000 & 0.87 (0.66) & 8 & 461 (300) & 895 (624) \\
        12000 & 0.43 (0.33) & 8 & 923 (600) & 2474 (1795)\\
        \hline
    \end{tabular}
    \caption*{Columns show the initial gas mass ($M_{gas}$), simulation duration from the formation of the first sink particle ($T_{end}$), final star formation efficiency (SFE), the final total mass in sink particles ($M_\star$), and the final number of sink particles produced ($N_\star$). The values in parenthesis are the corresponding column value at SFE=0.05.}
    \label{tab:simulation_summary}
\end{table}

\section{Method}
\label{sec:method}

\subsection{Simulation setup}
\label{sec:simulation_setup}

We use a locally developed version of the publicly available magnetohydrodynamics (MHD) Adaptive mesh refinement (AMR) code \texttt{Ramses} \citep{teyssier_cosmological_2002}, which is described in \citet{haugbolle_stellar_2018}. To explore how binarity is affected by the environment we have carried out six models of a star-forming region. These are an extension of the models discussed in \citet{haugbolle_stellar_2018} and we refer the reader to for more details about the methodology and numerical methods. The models have initial gas masses of $M_{gas}=$1500, 3000, 3750, 4500, 6000 and 12000 $\msun$ and a periodic box size of $L_{box} =4$ pc. The equation of state is isothermal assuming a 10 K gas. The initial conditions are homogeneous and contain a magnetic field of $7.2\mu\mathrm{G}$, initially aligned with the z-axis, corresponding to an average Alfvenic Mach number of 5.

To create a state that reflects the observed density-velocity-magnetic field relations we use continuous solenoidal random turbulent driving on the largest scale with an amplitude that results in a typical 3D velocity dispersion of 2 $\kms$. Initially, turbulence is driven for 20 dynamical timescales to erase any memory of the homogeneous initial conditions. Then gravity and sink particle formation are turned on while driving is maintained. The models are evolved until they reach a star formation efficiency (SFE) of at least 5\%. The SFE is defined as the fraction of the initial gas mass that is accreted into sink particles.

We use a root grid of 256$^3$ and add six levels of AMR refinement. The refinement strategy is based on the over-density. We refine the root grid when the density reaches a threshold density such that the Jeans length is resolved by 14.4 cells, and then refine the grid every time the density increases by a factor of four, keeping the minimum number of cells per Jeans length constant on all levels, except the highest level of refinement. The resolution on the highest level of refinement is $50\au$, and the density threshold for sink formation is $1.7\times,10^9$ cm$^{-3}$. Details of each simulation are summarised in \Cref{tab:simulation_summary}.

\subsection{Sink particle model}
\label{ssec:sink_model}
A sink particle model is implemented to capture star formation. A summary of the sink particle formation parameters is below. For a detailed description of the sink particle model, readers can refer to \cite{haugbolle_stellar_2018}. For a sink particle to form within our simulations, the collapsing gas must meet the following criteria:

\begin{enumerate}
\item The cell is on the highest level of refinement.
\item The density within a cell must be above the threshold number density of $\rho_{sink}= 1.7 \times 10^9$cm$^{-3}$. A typical value for $\rho_{sink}$ in the runs presented here is $10^5$-$10^6$ times the average density.
\item There must be a gravitational potential minimum in the cell where the sink particle will form.
\item The velocity field is converging at the cell.
\item No other sink particle exists within an exclusion radius. For our simulations, this is set to 8 cells or $400\au$.
\item The Jeans' length is resolved with at least two cells.
\end{enumerate}

This sink particle prescription is similar to others used in different MHD codes \citep{bate_modelling_1995, krumholz_embedding_2004, federrath_modeling_2010, gong_implementation_2013}.

When a sink particle forms, it initially has no mass, but mass is accreted from cells within the accretion radius ($r_{acc}$). In our simulations $r_{acc}$ is set to four cells. The accretion rate to the sink particles depends on the gas density and how well coupled a given gas cell is to the nearest sink particle; see \citet{haugbolle_stellar_2018} for details.

Protostellar jets and outflows carry away a portion of the mass and momentum that is accreted to the disc region \citep{tomisaka_evolution_2000, tomisaka_collapse_2002,federrath_modeling_2014, kuruwita_binary_2017}. However, with a minimum cell size of 50 AU, we only account for the infall of gas from the environment to the disc-star system, and the launching region of these outflows is not resolved. To account for mass loss via outflows, we use a reduced accretion efficiency: only 50 \% of the gas mass and momentum that is removed from the gas due to sink accretion is added to the sink particles while the other 50 \% is removed from the simulation. For example when reaching an SFE of 5\% only 95\% of the mass is available in gas and sinks. For low SFEs, the removal of gas does not impact the global gas reservoir but reduces on average the stellar masses by a factor 3 \citep{haugbolle_stellar_2018}.

Sink particle positions and velocities are updated using a leap-frog integrator. The smoothing length for the gravitational potential of the sinks is set to be $0.33\Delta x/ \approx 16.6\au$.

With the selected simulation parameters and sink particle model, our simulations resolve core fragmentation but not disc fragmentation. This choice is necessary to limit the already very high computational cost and makes it reasonable to use an isothermal equation of state. Furthermore, it is a perfect setup to study the formation and evolution of multiple star systems produced exclusively via core fragmentation and dynamical capture.

\section{Results and discussion}
\label{sec:results}

\begin{figure}
    \centerline{\includegraphics[width=0.9\linewidth]{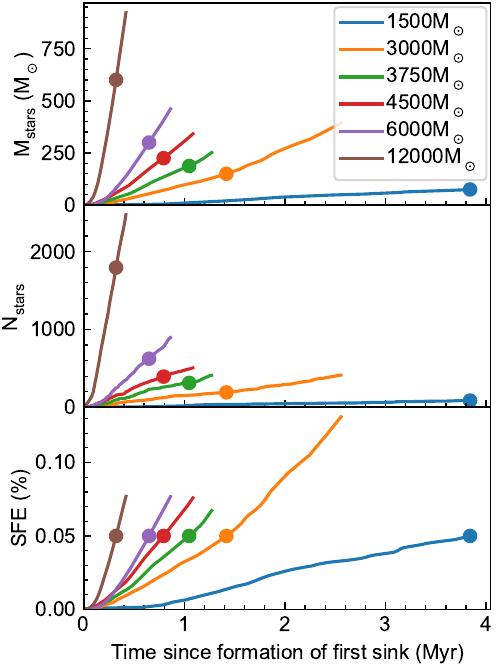}}
    \caption{Evolution of the total mass accreted in sink particles (\emph{Top}), number of sink particles (\emph{Middle}), and star formation efficiency (SFE, \emph{Bottom}) over time for all simulations. The circles indicate the value of a quantity when the simulation has SFE=0.05.}
    \label{fig:simulation_evolution}
\end{figure}

The evolution of the total accreted mass, number of stars, and SFE of all simulations are shown in \Cref{fig:simulation_evolution}, and values are summarised in \Cref{tab:simulation_summary}. \Cref{fig:simulation_evolution} shows that simulations with higher initial gas mass have a higher star formation rate \citep{padoan_star_2011, padoan_simple_2012, federrath_star_2012} and also a higher rate of stars formed \citep{haugbolle_stellar_2018} in accordance with theory. The high mass simulations also begin to produce sink particles earlier than the lower mass simulations, with the first sinks in the $12000$~M$_\odot$ simulation forming around $22$~Myr, while the $1500$~M$_\odot$ simulation begins forming its first sinks around $23$~Myr, as shown in \Cref{tab:simulation_summary}.

\begin{figure*}
    \centerline{\includegraphics[width=0.955\linewidth]{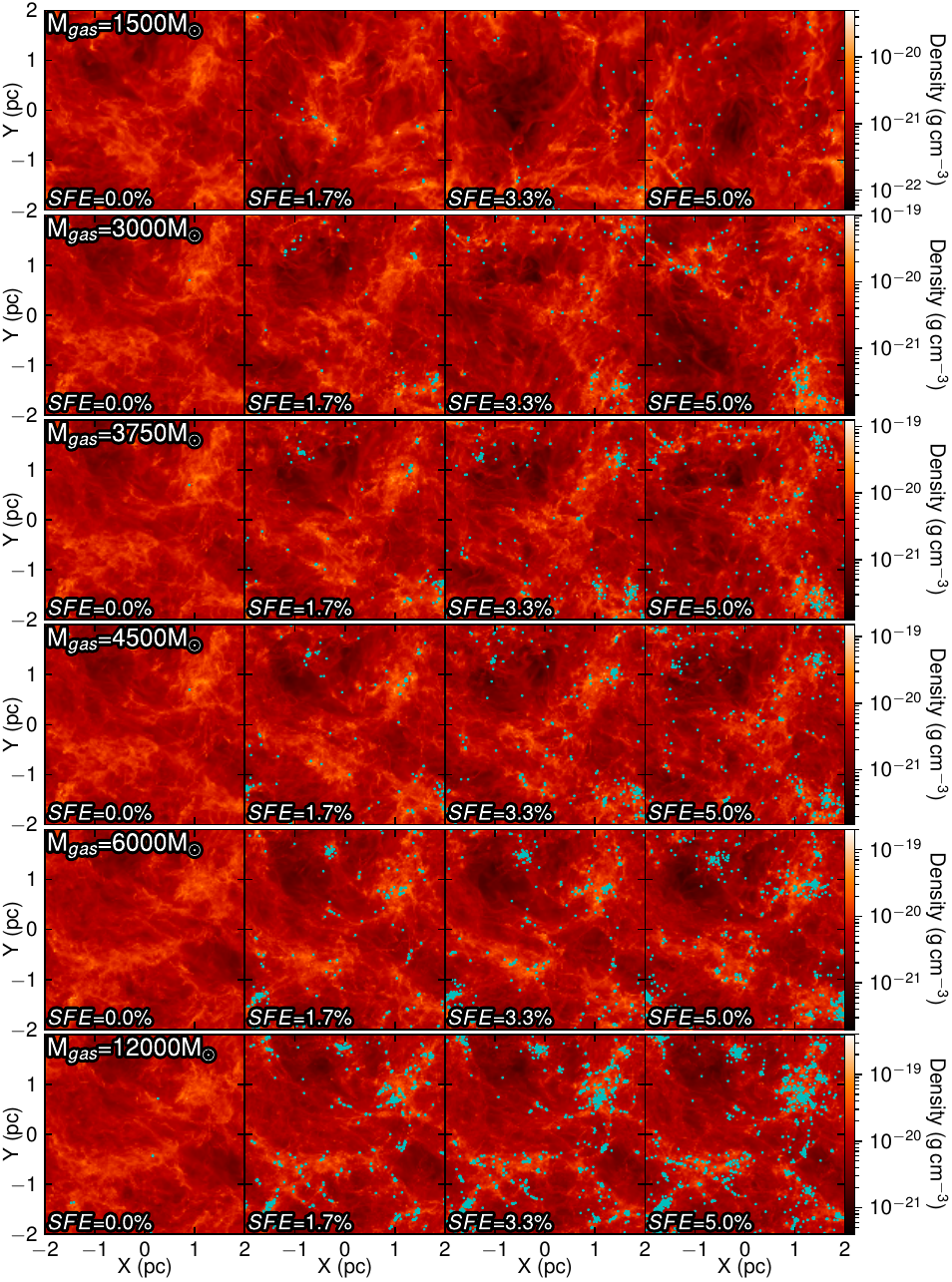}}
    \caption{Projections of the density over the entire computational domain from the formation of the first sink (\emph{Left column}) to SFE=0.05 (\emph{Right column}), for each simulation. The sink particles are annotated with blue particles. The colour bars are logarithmically centred on the mean gas density of the initial gas distribution.}
    \label{fig:simulation_projections}
\end{figure*}

Projections of the density over the entire computational domain for each simulation are shown in \Cref{fig:simulation_projections}. Each row is a simulation and each column shows a projection for SFE=0\%, 1.7\%, 3.3\%, and 5\% respectively. The colour bars are centred logarithmically on the initial mean gas density of each simulation (defined as $M_{gas}/4$pc$^3$) and have the same range. This was chosen to visualise degrees of gas concentration between simulations. The sink particles are annotated by blue dots. From the projections, we see that most sink particles reside in regions of high column density, which, in particular at early times, is close to their region of birth.

\subsection{Identifying multiple star systems and their formation pathways}
\label{ssec:formation_pathways}

As the sink particles accrete and interact, multiple star systems form and disintegrate. Multiple star systems were identified by:
\begin{enumerate}
    \item Finding the sink particle pairs that are below a separation threshold and gravitationally bound (where the $E_{potential} + E_{kinetic} = E_{total} < 0$). For our work, the maximum separation is set to $10\,000\au$.
    \item We then sort the pairs from lowest to highest separation.
    \item A new `particle' is saved with the position being the centre of mass of the bound sink particle pair and mass that is the total mass of the particle pair.
\end{enumerate}

\begin{figure*}
    \centerline{\includegraphics[width=0.79\linewidth]{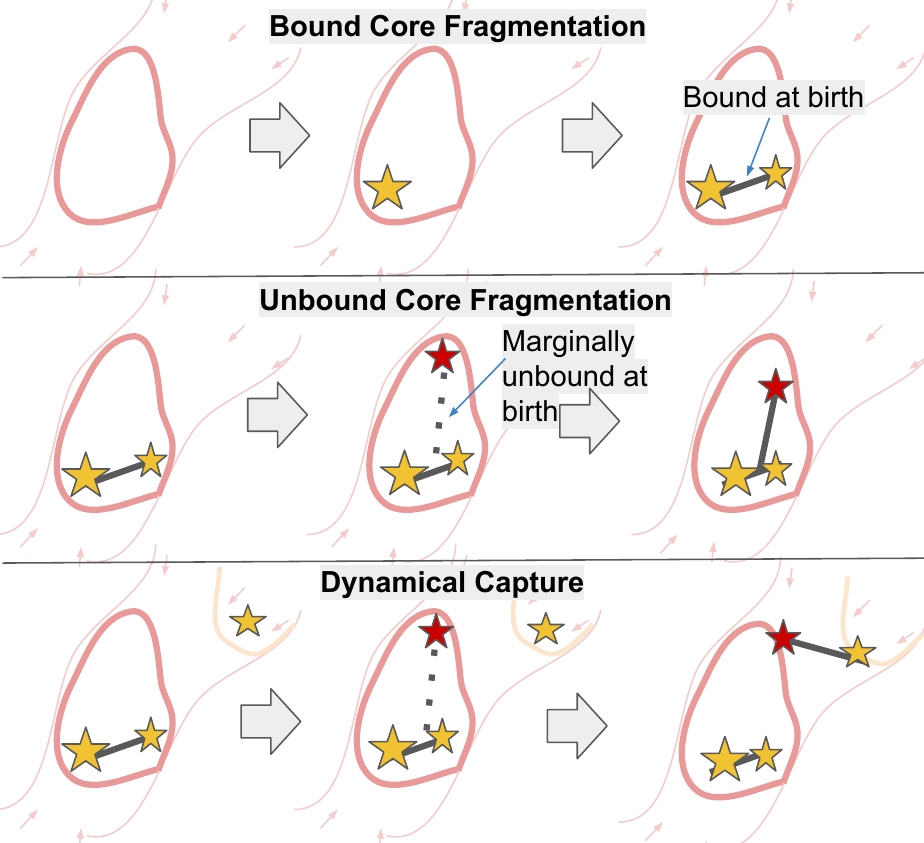}}
    \caption{Illustration of the three formation pathways that a system with a new sink ID is classified into: \emph{Top}: If a sink is formed and is gravitationally bound to another sink or stellar system at birth (marked with a solid line), this is bound core fragmentation, \emph{Middle}: If a sink is formed and is gravitationally unbound at birth, but later becomes bound to the sink or stellar system it was most bound to at birth (marked with a dashed line), this is unbound core fragmentation, and \emph{Bottom}: If a sink is formed and is gravitationally unbound at birth, and later becomes bound to a sink or stellar system that is different to the one it was most bound to at birth, this is dynamical capture.}
    \label{fig:formation_pathway_schematic}
\end{figure*}

\begin{figure*}
    \centerline{\includegraphics[width=0.92\linewidth]{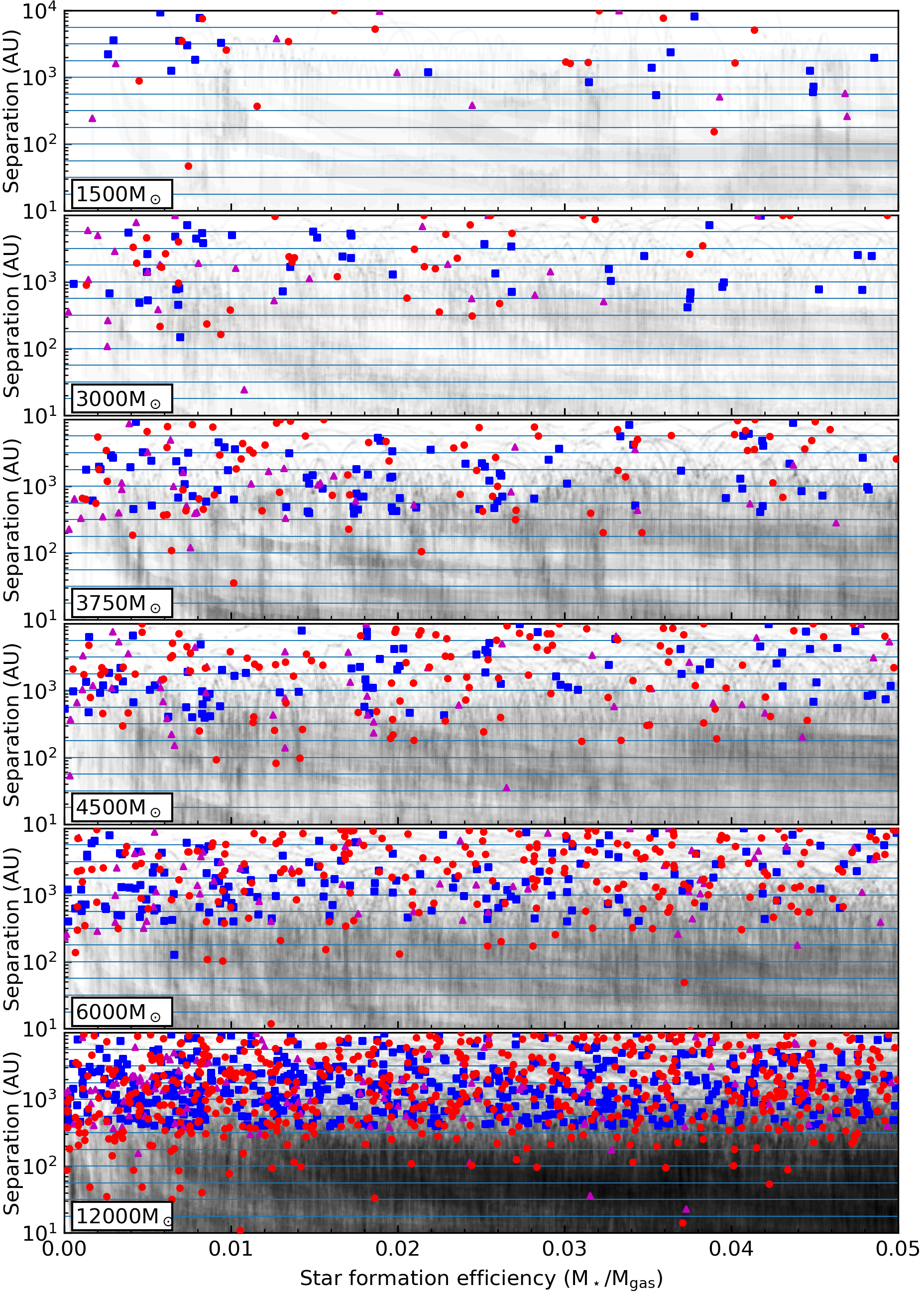}}
    \caption{Separation vs SFE of all systems formed in each simulation. The markers indicate the formation pathway of the system if it contains a new sink id. \emph{Blue boxes}: Bound core fragmentation, \emph{purple triangles}: unbound core fragmentation, \emph{red circles}: dynamical capture.}
    \label{fig:separation_vs_SFE}
\end{figure*}

This process is carried out recursively until all bound systems with up to six components are identified. Six components are chosen as the maximum number of stars in a system because higher multiples systems tend to be short-lived with very weakly bound other companions in a highly dynamic environment, such as close to the hub-like clusters in the simulations. Sextuples are also the largest systems identified in the observations by \citet{tobin_vla_2016}. When comparing our simulations with observations we change the approach for identifying multiple systems since three-dimensional distances and energies are not available for observations. This is discussed in \Cref{sec:obs_comparison}.

When iterating, to avoid double counting sink particles that are already in a bound system, only systems are used to calculate pairs. These are either an actual sink particle if it is single, or one of the systems that was created to represent a bound pair.

\begin{figure}
    \centerline{\includegraphics[width=0.95\linewidth]{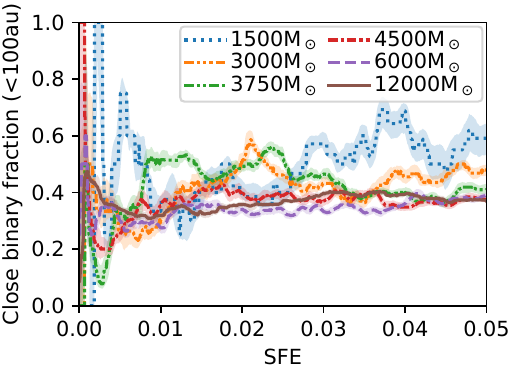}}
    \caption{Fraction of all separations (between stars/stars and centre of masses/between centre of masses) that are below $100\au$, over time for all simulations. The fraction was smoothed over a window of SFE$\pm5\times10^{-4}$. The solid line shows the median of this smoothing and the shaded regions show the $1\sigma$ variation of the smoothing.}
    \label{fig:close_sep_fraction}
\end{figure}

As highlighted in \Cref{ssec:sink_model} given the numerical resolution in our models, we can only capture core fragmentation. However, some sink particles can form isolated and unbound, but later become bound to another star. We use the birth conditions of a sink particle found in a binary or multiple to determine the pathway in which the system formed. We chronologically check bound systems, and if a new sink particle ID is found, we determine the system's formation pathway using the following criteria:
\begin{enumerate}
    \item {\bf Bound core fragmentation}: If a sink is formed bound to another sink or system of sinks, then it was formed via bound core fragmentation.
    \item {\bf Unbound core fragmentation}: If a sink is formed unbound, but the sink or systems of sinks it was most bound to at birth is in the current multiple system, then it was formed via unbound core fragmentation.
    \item {\bf Dynamical capture}: If a sink is formed unbound, and the sink or system of sinks it was most bound to at birth is not in the current multiple system, then it was formed via dynamical capture.
\end{enumerate}

These formation pathways are illustrated in \Cref{fig:formation_pathway_schematic}, and examples from the simulations are shown in the Appendix in \Cref{fig:pathway_appendix}. For systems with more than two stars, the formation pathway is determined by the newest sink particle ID ie. the most recently formed star. This classification only relies on the energy of the sink particles, and not that of the gas. Therefore, some of the systems found to be formed by unbound core fragmentation may in reality be bound, when making a detailed account of the gas distribution. We have chosen not to do this since it would entail relating different volume elements to different sink systems, which is not easily done in a unique manner and is beyond the scope of this article.

In \Cref{fig:separation_vs_SFE}, we plot the separations of all multiple systems for each simulation. For binaries, this is the true separation, but for higher-order multiples, this may be the separation between stars and the centre of mass of an inner system, or two centres of mass, depending on the system configuration. If a new sink particle ID is found in a system, we also annotate its formation pathways with a marker at the initial separation of the system. In \Cref{fig:separation_vs_SFE}, the blue boxes indicate systems that formed via bound core fragmentation, the purple triangles indicate systems that formed via unbound core fragmentation, and the red circles indicate systems that formed via dynamical capture. We see that all formation pathways occur throughout the simulations. Systems that have a small initial separation likely formed via dynamical captures when a stellar fly-by of a binary resulted in a three-body interaction that expels one of the original companions and captures the fly-by star resulting in a new binary system.

In \Cref{fig:separation_vs_SFE}, we see in each panel that many systems experience rapid orbital shrinkage, where the final separation of the systems is more than an order of magnitude smaller than the initial separation. In \Cref{fig:close_sep_fraction} we show the fraction of separations that are below $100\au$ over time for all simulations. In all simulations, the fraction initially varies significantly due to small number statistics. However, after approximately SFE=0.02, a gradient starts to appear with the lowest mass and highest mass simulations having 60\% and 40\% of separations below $100\au$, respectively. We investigate the origin of this gradient by looking at how clustered the star formation is, and orbital evolution in the following sections.

\begin{figure}
    \includegraphics[width=0.95\linewidth]{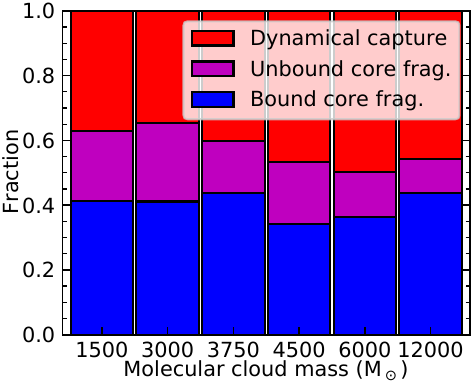}
    \caption{Fractions of systems formed via the pathways (illustrated in \Cref{fig:formation_pathway_schematic}) for each simulation: bound core fragmentation (blue), unbound core fragmentation (purple), and dynamical capture (red).}
    \label{fig:formation_pathway_frac}
\end{figure}

In \Cref{fig:formation_pathway_frac} we plot the fractions of systems that form via the different pathways for each simulation. The fractions are broadly consistent over all simulations. The fraction of systems that formed via bound core fragmentation varies between $35-45\%$, unbound core fragmentation varies between $10-25\%$, and dynamical capture varies between $35-50\%$.

The fraction of systems formed through unbound core fragmentation decreases with gas density, and hence stellar density. This is probably caused by the more dynamic environment seen in the higher-density models, leading to a lower time scale for encounters with other stars.

Overall, the fractions of that systems formed via core fragmentation (bound and unbound) are slightly higher in the lower mass simulations. This may be because the lower mass simulations require a larger over-density for protostellar cores to collapse, and this results in more systems forming via the core fragmentation pathways. This implies that star formation in lower mass GMC is more clustered than in higher mass one.

\subsection{Measuring clustering with two-point correlation function}
\label{ssec:clustering}

In this section we measure three-dimensional two-point correlation functions (TPCF) to gain insight into how clustered the sink particle distributions are in each simulation. A TPCF quantifies clustering by measuring all possible separations between points (for $N$ stars, the number of separations is $N(N-1)/2$)) and binning the separations into logarithmic separation bins ($r$). This is calculated for the observed data points ($P_{\mathrm{DD}}(r)$) and for randomly generated points that are uniformly distributed ($P_{\mathrm{RR}}(r)$). These two distributions are normalised using the number of data points and randomly generated stars to obtain:

\begin{equation}
    \mathrm{DD}(r) = \frac{P_{\mathrm{DD}}(r)}{N_{\mathrm{D}}(N_{\mathrm{D}} - 1)},
    \mathrm{RR}(r) = \frac{P_{\mathrm{RR}}(r)}{N_{\mathrm{R}}(N_{\mathrm{R}} - 1)},
\end{equation}

\noindent where $N_{\mathrm{D}}$ and $N_{\mathrm{R}}$ is the number of data points and randomly generated points respectively. The basic TPCF \citep{peebles_large-scale_1980} is derived by dividing the normalised distribution of the data (DD$(r)$) by the normalised distribution of the random points (RR$(r)$), that is:

\begin{equation}
    1+\omega(r) = \frac{\mathrm{DD}(r)}{\mathrm{RR}(r)},
    \label{eqn:TPCF}
\end{equation}

If the data are clustered, the resulting TPCF will have values $\gg1$ in smaller separations and typically has a power law function as a function of separation with a negative slope.

To calculate a TPCF, an underlying distribution must be derived, for comparison. Given the periodic boundary conditions of our simulation, the largest separation two sinks can have is $\sqrt{3(L/2)^2}\sim$3.5$\,$pc which is equivalent to $\sim$7$\times 10^5\,$au. With this largest separation and the smoothing length of 16.6\,au, 10 separation bins spaced logarithmically between $10-10^6\,$au were selected to calculate the TPCF. To derive the underlying distribution, $5\times10^5$ positions were randomly generated within the simulation domain, and separations were calculated. A large number of randomly generated positions was necessary to populate the smallest separation bin. 100 instances of $5\times10^5$ randomly generated positions were performed, and the final underlying distribution ($P_{\mathrm{RR}}(r)$) was taken to be the mean of these 100 instances, and the error is the standard deviation in a separation bin over the 100 instances.

For the simulations, the distribution ($P_{\mathrm{DD}}(r)$) is derived by calculating all separations between stars and binning the separations. The error of this distribution is taken to be the Poisson noise, because of the low number of stars per separation in the simulation, i.e. $\sigma_{P_{\mathrm{DD}}(r)}=\sqrt{N}$.

\begin{figure}
    \includegraphics[width=0.95\linewidth]{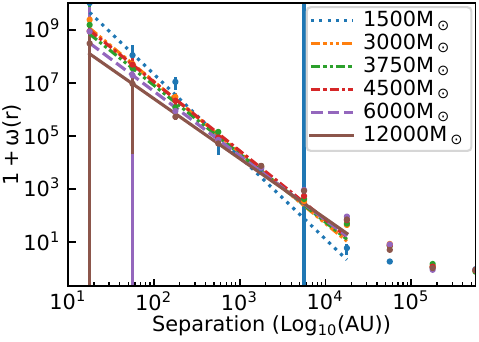}
    \caption{Two point correlation functions (TPCF) of all simulations at SFE=0.05. The error bars on the points are the standard deviation of the TPCF values calculated against the 100 generated instances of a uniform distribution. The lines are the power law fits derived from the TPCFs up to $1 - 3\times 10^4\au$ separation bin.}
    \label{fig:TPCF_SFE_5}
\end{figure}

The TPCFs for all simulations at SFE=0.05 are shown in \Cref{fig:TPCF_SFE_5}. The error bars for the TCPF are derived by summing the relative errors of DD$(r)$ and RR$(r)$. After calculating the TPCF a power-law is fit to the function, with a steeper slope indicating stronger clustering. Empty bins are masked to avoid poor power-law fits. At SFE $=0.05$, the lower mass simulations show stronger clustering, however, this is only at one snapshot in time.

\begin{figure}
    \includegraphics[width=\linewidth]{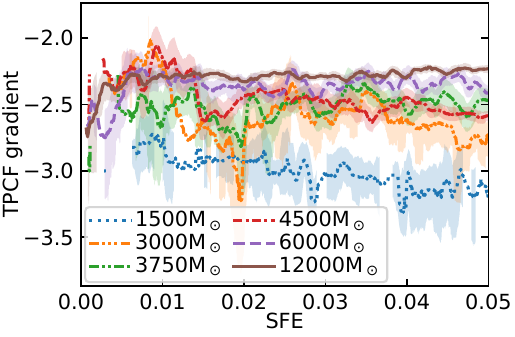}
    \caption{Gradient of two-point correlation function (TPCF) over each simulation. The smoothing was carried out in the same way as \Cref{fig:close_sep_fraction}, however, the shaded region shows the sum of the 1$\sigma$ error on the power-law index and the 1$\sigma$ variation in a smoothing window.}
    \label{fig:TPCF_grad}
\end{figure}

To understand how the clustering evolves over the course of the simulations, in \Cref{fig:TPCF_grad} we plot fitted power-law indices for each simulation as a function of SFE. The shaded region shows the sum of the $1\sigma$ error on the power-law index plus the $1\sigma$ variation in a smoothing window. Initially, fitting a gradient is not possible due to small number statistics leading to many empty separation bins. The $M_{gas}=1500\,$M$_\odot$ simulation has a stronger and increasing clustering throughout the simulation. Conversely, the higher-density simulations typically have shallower gradients, with little time evolution. The difference in clustering is likely due to the low-density simulation producing no stellar clusters and mostly binaries, with very few high-multiplicity systems, giving a more peaked TPCF. This is because the virial parameter is so high that stars are only formed a few at a time at density peaks with an exceptionally high over-density, leading to a very slow and stochastic star formation.

Observationally, the two-point correlation functions in the nearby Taurus, $\rho$ Oph, and Orion regions are found to be a power law with exponents in the range $-2 < \beta < -1$ \citep{gomez_spatial_1993, larson_star_1995, gomez_head_1998}, which compares well to our results, given that exponent should be increased by one when projecting from three to two dimensions.

\subsection{System formation scales}

\begin{figure*}
    \centerline{\includegraphics[width=0.7\linewidth]{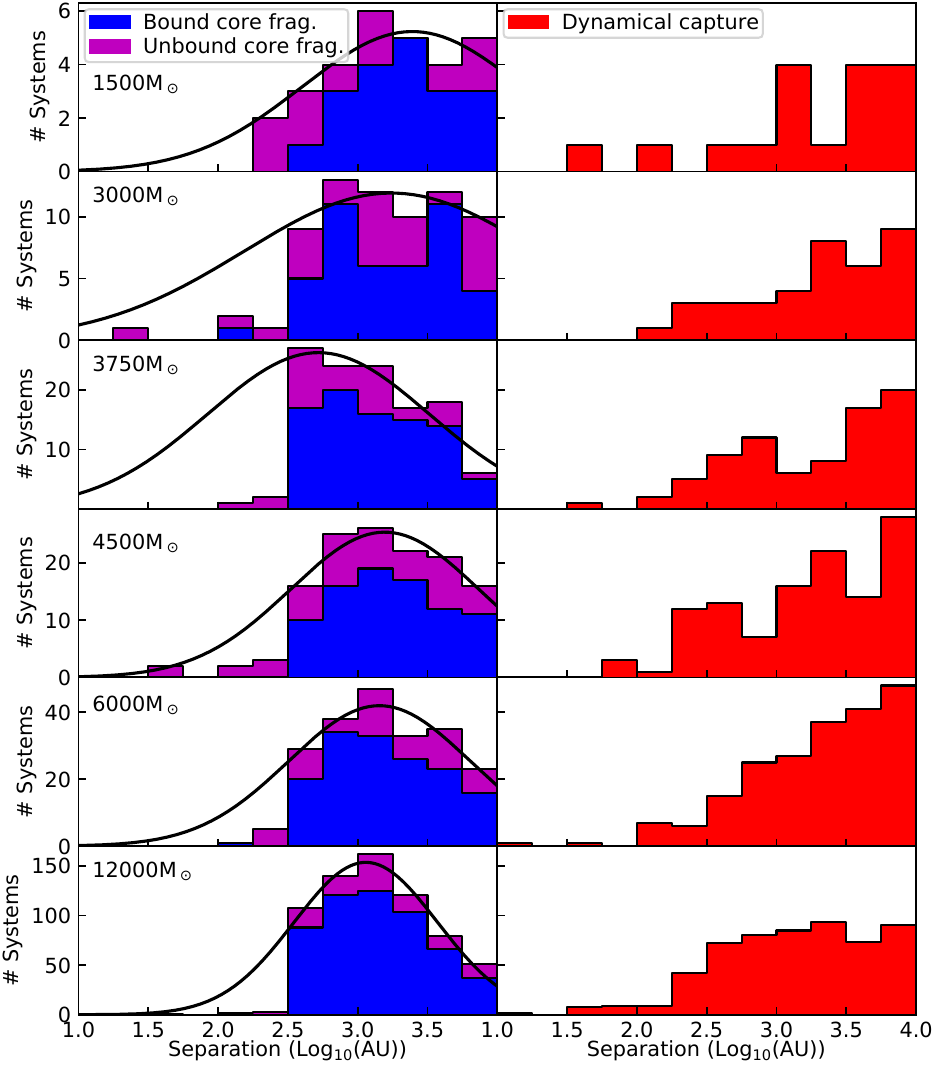}}
    \caption{Histograms of the initial separations of systems with new sink IDs for each simulation, for the three different formation pathways: \emph{Blue}: bound core fragmentation, \emph{purple}: unbound core fragmentation, and \emph{red}: Dynamical capture. The separations were binned in log(Separation), and the histogram for unbound core fragmentation is stacked on top of the bound core fragmentation histogram. The black line is the fitted Gaussian to the core fragmentation distribution.}
    \label{fig:initial_separations}
\end{figure*}

The star-forming environment may affect the initial separation of systems formed via core fragmentation pathways, which might give insight into a fragmentation scale. In \Cref{fig:initial_separations} we plot the histogram of the initial separations of the systems formed via the three different formation pathways for each simulation pathway. For the core fragmentation pathways, the histograms are stacked in the left column, while the initial separation of dynamical capture is shown on the right column of \Cref{fig:initial_separations}.

\begin{figure}
    \includegraphics[width=\linewidth]{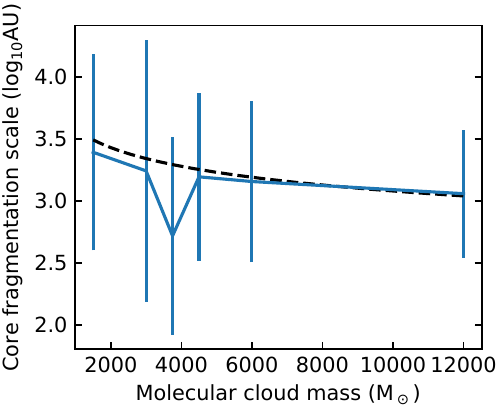}
    \caption{Mean fragmentation scale as a function of the GMC mass. The curve shows the peak of Gaussian distributions fitted to the (stacked) histogram of initial separations of systems formed via core fragmentation pathways plotted in \Cref{fig:initial_separations}. The error bars are the width of the Gaussian distributions and not the error on the peaks. The dashed line shows the theoretical prediction from assuming that cores on average can be described as critical Bonnor-Ebert spheres.}
    \label{fig:frag_scale}
\end{figure}

We fitted Gaussian distributions to the sum of the histograms for systems formed via bound and unbound core fragmentation to characterise a typical bound core fragmentation scale, $d_{core}$. The means and widths of these distributions are plotted against GMC mass in \Cref{fig:frag_scale}. The fragmentation scale appears to be smaller in denser environments, however, the typical fragmentation scale over all simulations appears to be around $10^{3\mbox{--}3.5}\au$ which is consistent with observations that find a peak of around $\sim3000\au$ in the separation distribution of protostars \citep{pokhrel_hierarchical_2018}. 

If we assume that the cores on average can be described as critical mass Bonnor-Ebert spheres, the external pressure on the cores will increase with increasing cloud mass, leading to the core fragmentation scale depending inversely proportional to the square root of the cloud mass, indicated by the dashed line in \Cref{fig:frag_scale}.

The normalisation is set by assuming that the typical scale at which binaries form is similar to the radius of a critical Bonnor-Ebert core:
\begin{equation}
d_{core} = \epsilon_{core}\,R_{BE} = 0.485 \,\epsilon_{core}\,\frac{c_s^2}{G^{1/2} P_{ext}^{1/2}},
\end{equation}
where $\epsilon_{core}$ encapsulates that a binary companion will not form at the very edge of a pre-stellar core (we use $\epsilon_{core} = 0.55$ in the figure), and
$P_{ext}$ is the confining pressure of the Bonnor-Ebert sphere, which can be estimated as \citep{haugbolle_stellar_2018}:
\begin{equation}
P_{ext} = (1 + \mathcal{M}_s^2) P_{th}
= (1 + \mathcal{M}_s^2) \, c_s^2 \frac{M_{gas}}{L_{box}}.
\end{equation}

When looking at the initial separations of systems formed via dynamical capture (shown on the right column of  \Cref{fig:initial_separations}), we see that for all GMC masses, the histogram increases with larger initial separations. For simulation $12000\msun$ the initial separation histogram plateaus beyond $1000\au$. This plateau is likely due to the higher stellar density which creates a situation where it is less probable to be captured at a large distance from a stellar system because other systems will be nearer.

\subsection{Orbital evolution of the different pathways}
Simulations of binary star formation via protostellar core fragmentation often show that the binaries undergo significant orbital migration. There are two classes of mechanisms that can efficiently provide migration. Dynamical friction \citep{chandrasekhar_dynamical_1943,ostriker_dynamical_1999}, slow-down due to accretion of material with lower angular momentum, and the initial high-eccentricity configuration of newly formed cores is a combination of reasons why naturally binaries with separations below $\sim1000\au$ suffer rapid inspiral. Prestellar cores form in general along filaments, which are kinematically cold in the direction transverse to the filament. Therefore, binaries formed through core fragmentation will initially have an angular momentum, much lower than what is required for circular motion and launch on highly eccentric orbits. This low angular momentum leads to rapid inspiral of the young binary system. Systems formed via dynamical capture must instead go through a state where they are marginally unbound and then become bound. When these systems first form, their angular momentum is therefore high, and thus, these systems are not expected to undergo fast orbital evolution. An exception is when dynamical capture is the end product of a three-body interaction, which can lead to an interchange of the stellar companions and hardening of the binary system \citep{reipurth_disintegrating_2000}.

To investigate inspiral, we measure the rate of change of the semi-major axis in the $1000$ and $10\,000\yr$ after the first periastron of a system. The semi-major axis is calculated at all time steps using:
\begin{equation}
    a = -\frac{G\,M_{tot}}{2\epsilon_{orb}}\,,
    \label{eqn:semimajor_axis}
\end{equation}
where $M_{tot}$ is the total mass of the system, and $\epsilon_{orb}$ is the specific orbital energy of the system. $\epsilon_{orb}$ is calculated by summing the specific kinetic and potential energies of the system.

\begin{figure}
    \includegraphics[width=\linewidth]{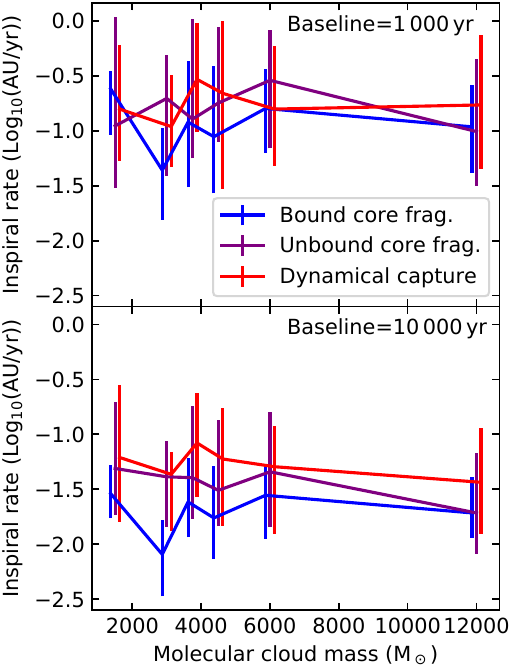}
    \caption{Median inspiral rate of systems that formed via the different formation pathways (see \Cref{fig:formation_pathway_schematic}) measured over the $1000\yr$ (\emph{Top}) and $10\,000\yr$ (\emph{Bottom}) after the first periastron of the system vs initial gas mass of each simulation. The error bars show the standard deviation of the measured inspiral rates.}
    \label{fig:inspiral}
\end{figure}

The median inspiral rate for systems of different formation pathways is plotted against simulation gas mass in \Cref{fig:inspiral}. The error bars indicate the standard deviation of the measured inspiral rates. For both baselines, the inspiral rates are independent of the average density of the star-forming environment. This may be because forming systems are gravitationally decoupled from the large-scale environment and typical densities and time scales that ultimately sets the in-fall rates are set by gravity.

We calculated the median inspiral rate indices over all simulations and finds that the indices are generally higher ($-0.80^{+0.60}_{-0.57}$) in the $1000\yr$ baseline compared to the $10\,000\yr$ baseline ($-1.50^{+0.55}_{-0.43}$) for all pathways. The error bars of the quoted indices are the $1\sigma$ variation over all simulations. This variation is not surprising because we expect the inspiral rate to be the greatest soon after the formation of the system, and then decrease at later stages.

For the $1000\yr$ baseline we do not see any significant variation between formation pathways. For the $10\,000\yr$ baseline, we see that the median inspiral rate of systems formed via dynamical capture is marginally higher ($-1.32^{+0.47}_{-0.54}$) than bound core fragmentation ($-1.68^{+0.36}_{-0.32}$) over all simulations.

The resulting inspiral rates measured from the $10\,000\yr$ baseline may seem contrary to the previously stated hypothesis, but this is a result of the typical initial separation of dynamical capture systems being larger than for the core fragmentation systems (see Fig.~\ref{fig:initial_separations}). Overall, we see that many systems experience significant orbital evolution within the early stages of their lifetime.

\section{Comparison with observations}
\label{sec:obs_comparison}

We now aim to compare the multiplicity statistics produced from these simulations with observations of protostellar multiplicity. In particular, we compare our results with that of \citet{tobin_vla_2016} of the Perseus star-forming region using the revised statistical method described in \cite{tobin_vlaalma_2022}. \citet{tobin_vla_2016} observed protostars in this region and measured how the companion frequency (CF) evolved with the separation of the components. The companion frequency as defined by \citet{reipurth_visual_1993} is the average number of companions a star has and is given by:

\begin{equation}
    CF = \frac{B+2T+3Q+...}{S+B+T+Q+...},
\end{equation}

\noindent where $B, T, Q, ...$ are the number of single, binary, triple, quadruple, and higher order systems respectively.

\citet{tobin_vla_2016} observed a bimodal distribution in the CF versus separation, with peaks at $\sim$75$\au$ and $\sim$3000$\au$ for Class 0 and I objects. The peak at $75\au$ was attributed to binaries formed via disc fragmentation while the peak at $3000\au$ was attributed to systems formed via core fragmentation. However, we have shown in the previous sections that multiple systems that formed on core fragmentation scales often experience significant orbital evolution, down to separations $\le75\au$. We aim to determine what the contribution of core fragmentation is to the observed bimodal distribution.

\subsection{The Perseus star-forming region and finding the best simulation for comparison}
\label{ssec:best_sim}

The Perseus star-forming region consists of multiple star-forming clusters. \citet{arce_complete_2010} presents a comprehensive analysis of the mass, volume, turbulence, and star formation efficiency (SFE) of various star-forming regions in Perseus. From \citet{arce_complete_2010} the estimated current mass, including gas mass, mass in outflows, and young stellar objects, in the L1448, NGC1333, B1-Ridge, B1, IC348, and B5 star-forming regions is approximately $3220\msun$. The estimated total volume of all star-forming regions is $53.5\pc^3$, which is slightly smaller than the $4^3=64\pc^3$ volume of the simulations. Based on these mass and volume estimates, the $M_{gas} = 3000\msun$ or $3750\msun$ simulations are best for the comparison.

To further refine which simulation is best for comparison, we calculate which simulation produces a similar number of sink particles with luminosities that would be observable by the observations of \citet{tobin_vla_2016}.

The bimodal distribution found by \citet{tobin_vla_2016} is only prominent in Class 0 and I objects, and it is not seen in Class II objects. Therefore, for this comparison, we aim to target sink particles that would be classified as Class 0/I. Accretion in protostars is generally higher in the early protostellar stages than in the later stages, so an accretion limit of $10^{-7}\,\mathrm{M}_\odot\,\mathrm{yr}^{-1}$ is applied to select sinks that are likely to be in the Class0/I stages. Observationally, Class I objects have observed accretion rates from $\sim$10$^{-9}$ to $10^{-6}\,\mathrm{M}_\odot\,\mathrm{yr}^{-1}$, however, Class II are not observed to have accretion rates above $10^{-7}\,\mathrm{M}_\odot\,\mathrm{yr}^{-1}$ \citep{fiorellino_KMOS_2021}. With the selected accretion limit we may miss some Class I objects with a low accretion rate, but avoid selecting Class II objects.

The number of Class 0/I objects observed by \citet{tobin_vla_2016} was $92$ stars, with luminosities between $0.1$ and $55\,\mathrm{L}_\odot$. In \citet{tobin_vlaalma_2022}, the upper limit of the Perseus observations is stated to be approximated $120\,\mathrm{L}_\odot$. To find the best simulation for comparison with these observations, we calculated the number of sink particles in a simulation that would be observable. We used a lower luminosity limit of $0.1\,\mathrm{L}_\odot$ and upper limits of both the highest luminosity object ($55\,\mathrm{L}_\odot$) and the theoretical maximum luminosity ($120\,\mathrm{L}_\odot$). We assume that the luminosity is dominated by the accretion luminosity. To calculate the accretion luminosity we use:

\begin{equation}
    L_{acc} = f_{acc}\frac{GM\dot M}{R_\star}
    \label{eqn:Accretion_L}
\end{equation}

\begin{figure}
    \includegraphics[width=\linewidth]{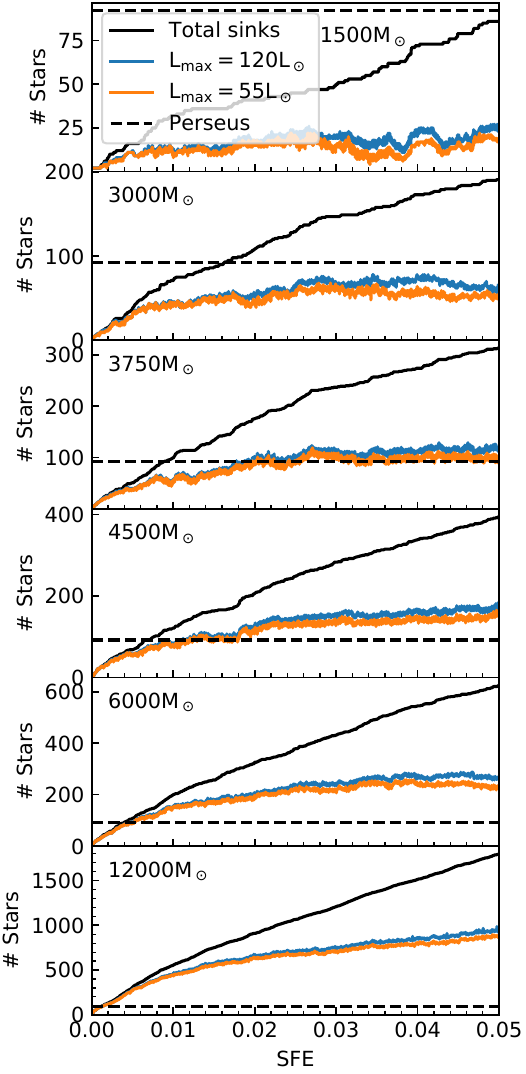}
    \caption{Total number of sink particles (black) and the number of visible sink particles for the maximum luminosity of $120\,\mathrm{L}_\odot$ (blue), and $55\,\mathrm{L}_\odot$ (orange) as calculated by \Cref{eqn:Accretion_L} over SFE for each simulation. The horizontal dashed line indicates the number of Class 0/I objects observed by \citet{tobin_vla_2016} (i.e. 92 stars).}
    \label{fig:visible_stars}
\end{figure}

\noindent where $\dot M$ is the mass accretion rate for the sink particle, $R_\star$ is the radius of the protostar, and $f_{acc}$ is the fraction of potential energy from accretion that is converted into radiation. For these calculations, we take $R_\star=2\mathrm{R}_\odot$ for all protostars and $f_{acc}=0.5$ (1.e. 50\% of potential energy is converted to radiation).

The evolution of the number of visible stars over time is shown in \Cref{fig:visible_stars}. We see that for all simulations, before SFE $\sim0.005$ essentially all sink particles are visible. However, at later times, the total number of sink particles and visible sink particles diverges. This is because older sink particles may have stopped accreting or have very low accretion rates, but newer sinks are forming which have higher accretion rates. In all simulations, a steady state in the number of visible sinks is established after SFE $\sim0.01$. Based on the steady-state visible star numbers and the number of objects observed by \citet{tobin_vla_2016} (annotated by the horizontal dashed line in \Cref{fig:visible_stars}) the $M_{gas}=3750\msun$ simulation produces the closest number of visible stars to the observations. Therefore, for further comparisons with observations, this simulation will be used.

\subsection{Calculating characteristic SFE of Perseus}

Observations capture the star formation evolution at a particular time, while simulations evolve over time. To make an appropriate comparison with observation, we need to find the best time for the simulations to compare with observations. To do this, we estimate a characteristic star formation efficiency of the observed star-forming regions in Perseus.

\citet{arce_complete_2010} provides right ascension and declination boundaries for the L1448, NGC1333, B1-Ridge, B1, IC348, and B5 star-forming regions and observational estimates for the star formation efficiency in these regions. Based on the boundaries of these regions, we find how many of the Class 0 and I objects from \citet{tobin_vla_2016} are in each region. We find that 12 stars are in L1448, 39 stars are in NGC1333, 4 stars are in B1-Ridge, 10 stars are in B1, 13 are in IC348, 2 are in B5, and 12 are unclassified. We re-classified the unclassified stars based on which star-forming region they are closest to on the sky, and this adds 11 stars to B1-Ridge and 1 star to B1.

We then calculate a weighted average of the star-forming efficiency based on the number of stars in each region and the observed SFE. Based on \citet{arce_complete_2010}, the SFE of each region is 1.5\% for L1448, 4.9\% for NGC1333, 2.4\% for B1-Ridge, 2.1\% for B1, 9\% for IC348 and 0.4\% for B5. From this, the characteristic SFE we retrieve is $\sim4.2\%$.

\begin{figure*}
    \includegraphics[width=\linewidth]{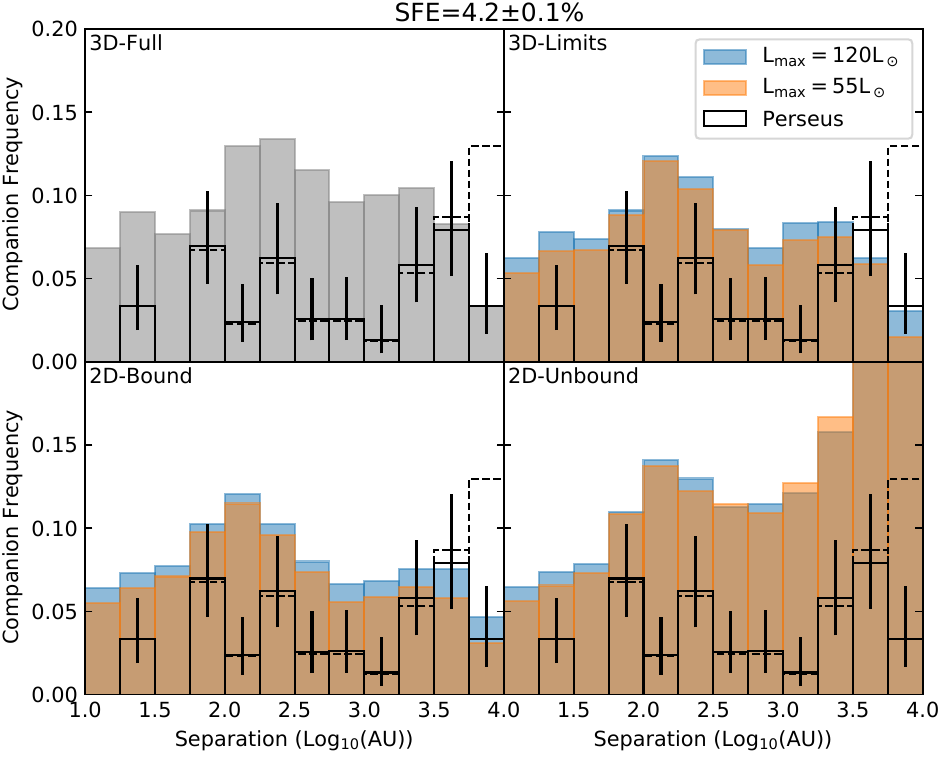}
    \caption{Median companion frequency vs separation at star formation efficiency of 4.2\%, averaged over an SFE $\pm0.1\%$ window, for the $M_{gas}=3750\msun$ simulation. The solid histograms in each panel show the resulting CF vs. Separation histogram for the different settings summarised in \Cref{tab:comparison_summary}. For settings with the luminosity limits, the blue and orange show the results with an upper limit of $55$ and $120\,\mathrm{L}_\odot$ respectively. The black dashed histogram shows the observed distribution found by \citet{tobin_vlaalma_2022} for all class 0/I objects, and the solid black line is the histogram derived from pairs with boundness likelihood $>0.68$. The error bars are calculated using the binomial statistics described by equation 3 in \citet{tobin_vlaalma_2022}.}
    \label{fig:CF_hist_42}
\end{figure*}

\subsection{Data processing to make CF versus separation histogram.}

To accurately carry out a comparison with observations, we must process our data the same way the observations were processed. While the true sink particle separations and separations between centres of mass for sub-systems were used to find bound systems previously, to be consistent with \citet{tobin_vlaalma_2022}, we also calculate the separations between mid-points of components and sub-systems. Readers can refer to Figure 1 of \citet{tobin_vlaalma_2022} for details on measuring separations in a multiple star system.

For some of the comparisons with observations, we save a projected midpoint-separation, for example, save the true separations projected onto the xy-plane. We also vary whether accretion and luminosity limits are applied to be consistent with \citet{tobin_vla_2016}, and whether to count unbound pairs (pairs with projected separation $<10,000\au$). The various analysis settings are summarised in \Cref{tab:comparison_summary}. After all the multiple systems are found based on the settings and limitations applied, we then process systems to determine how the companion frequency evolves with projected separation.

\begin{table}
    \centering
    \begin{tabular}{|l|c|c|c|}
        \hline
        Name & Projection & Bound & $\Dot{M}$ and $L_{acc}$\\
         & into 2D & Check & Limits\\
        \hline
        3D-Full & False & True & False\\
        3D-Limits & False & True & True\\
        2D-Bound & True & True & True\\
        2D-Unbound & True & False & True\\
        \hline
    \end{tabular}
    \caption{Settings used for CF vs. separation histogram. \emph{Name}: of set up, \emph{Projection into 2D}: projects midpoint separation onto the xy-plane, \emph{Bound Check}: only counts gravitationally bound pairs, \emph{Limits}: only counts `visible' stars based on accretion and luminosity limits (see \Cref{ssec:best_sim}).}
    \label{tab:comparison_summary}
\end{table}

We create 12 separation bins logarithmically spaced from $10^1 - 10^4\au$, in the same fashion of \citet{tobin_triple_2016} and \citet{tobin_vlaalma_2022}. Within each bin, if a multiple system has components with separations smaller than the lower bound of the bin, it is considered a single star, and if there are components with projected separations larger than the upper bound then that system is separated into smaller systems.

It is not appropriate to compare the observations with a single time step in the simulations because accretion rates and luminosity can vary between time steps. Therefore, the CF versus separation histogram was integrated over an SFE $\pm0.1\%$ window, and the median value in each bin is used to produce the resulting histogram.

For comparison with observations, we retrieved the observations for Perseus from \citet{tobin_vlaalma_2022}. We removed objects that were not Class 0/I and processed the data to create our histogram. We use the entire data set to produce the dashed histogram shown in \Cref{fig:CF_hist_42}. To approximate the correction for unbound pairs observed by \citet{tobin_vlaalma_2022}, we recalculate the histogram only for separation with boundness likelihoods $>0.68$ (Refer to Table 4 of \citet{tobin_vlaalma_2022}). This produces the solid line histogram shown in \Cref{fig:CF_hist_42}. The error bars are calculated using the binomial statistics described by equation 3 in \citet{tobin_vlaalma_2022}.

The results for SFE=4.2\% is shown in \Cref{fig:CF_hist_42}. For the 3D-Full settings, the true CF distribution of all bound systems appears to be relatively uniform at CF=0.1. However, the CF drops significantly in the last separation bin, indicating that there are fewer true bound systems at these large separations. When observational limits are imposed in 3D-Limits, we observe an overall reduction in CF in all bins. We also see a bimodal distribution appear with peaks at $\sim150\au$ and $\sim2000\au$. When projected onto 2D in 2D-Bound, the peak at $\sim150\au$ is still seen, but the larger peak is smeared out. Projecting separations into 2D will affect larger separations more than smaller separations, therefore it is expected that the CF histogram would be more affected at larger separations. When unbound pairs are counted in the 2D-Unbound analysis we see that the CF increases significantly above projected separations of $>100\au$.

\subsubsection{YSO density and contamination from chance alignments}
\label{ssec:yso_dens}

\citet{tobin_vlaalma_2022} correct their observations for possible chance alignments. The likelihood of a chance alignment is dependent on the observed young stellar object (YSO) density. Based on our 2D-Bound and 2D-Unbound analysis, \citet{tobin_vlaalma_2022} appears to have sufficiently filtered out chance alignments which may contaminate their results.

Our simulation produced significantly more unbound pairs than \citet{tobin_triple_2016}, and we hypothesise that the chosen simulation has a significantly higher stellar density than Perseus. To investigate this we calculate the YSO density distribution for our simulation at different stages of evolution.

The YSO density is calculated using the same method described by \citet{tobin_vlaalma_2022}. The YSO density around the sink particle is calculated using:

\begin{equation}
    \Sigma_{\mathrm{YSO}} = \frac{10}{\pi r_{11}^2}
    \label{eqn:yso_dens}
\end{equation}
 
\noindent where $r_{11}$ is the separation to the eleventh nearest visible neighbour.

\begin{figure}
    \includegraphics[width=\linewidth]{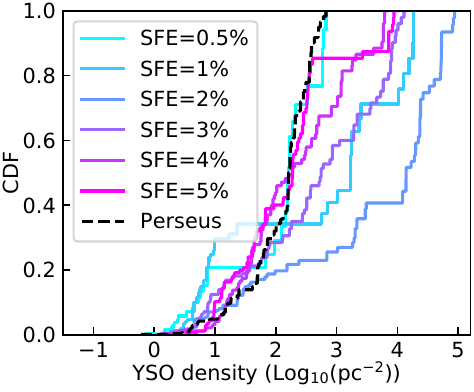}
    \caption{YSO density CDF for our simulations at different SFE (blue to pink gradient). The measured YSO density for \citet{tobin_vla_2016, tobin_vlaalma_2022} is shown by the dashed black line.}
    \label{fig:YSOdens}
\end{figure}

We measure the YSO density at SFE = 0.5\%, 1\%, 2\%, 3\%, 4\%, and 5\%. The resulting cumulative distributions are plotted in \Cref{fig:YSOdens} against the observed YSO density of Perseus for the Class 0/I objects. At SFE $>0.5\%$, the YSO density measured in the simulations is larger than Perseus. In the simulations, we observe that the maximum YSO density increases up to SFE=2\% and then decreases. This may reflect the initial burst of star formation in clusters, and then the later dispersal of these clusters.

The higher YSO density in the simulations contributes to the significantly higher CF measured with the 2D-Unbound setup in \Cref{fig:CF_hist_42}. The median YSO density for both Perseus and our simulation at SFE=4\% is approximately 100pc$^{-2}$, however, the simulation has an extended tail at higher densities. The highest YSO density measured among the Perseus objects is $10^{2.8}$pc$^{-2}$, while the highest YSO density in the simulation at SFE=4\% is $10^{3.8}$pc$^{-2}$, an order of magnitude greater. Therefore around some objects, we may expect up to 10 times more unbound pairs, contributing to the significant excess in companion frequency seen in \Cref{fig:CF_hist_42}.

Despite the significant increase in CF when counting unbound pairs, a peak in CF is still visible in this setup at $\sim$200$\au$.

\subsubsection{Finding best fit using two-sample KS test}

While we find a bimodal distribution at the calculated characteristic SFE, we aim to determine if there is another time in the simulation where a stronger bimodal distribution is found. We carry out a two-sample Kolmogorov–Smirnov (KS) test \citep{massey_kolmogorov-smirnov_1951} because it is a non-parametric comparison of the shape of two distributions. While we show the result of the bimodal distribution found in Perseus, we are not concerned by the values of the resulting CF versus separation histogram, but more specifically the shape.

To calculate the KS statistic, we:
\begin{enumerate}
    \item Order all the separations within systems observed in Perseus, and then calculate a cumulative distribution function (CDF). This distribution is normalised by dividing the CDF by the sum of all values (such that the bounds are 0 and 1)
    \item For each time in the simulation, we used the 2D-Bound analysis to find the visible stars. All separations in all bound systems are sorted to produce a normalised CDF for the simulated systems.
    \item Calculate the KS statistic, which is the largest vertical separation between the two normalised CDFs. We used the \texttt{scipy} package, which has the \texttt{scipy.stats.ks\_2samp} function to calculate the values.
\end{enumerate}

\begin{figure}
    \includegraphics[width=\linewidth]{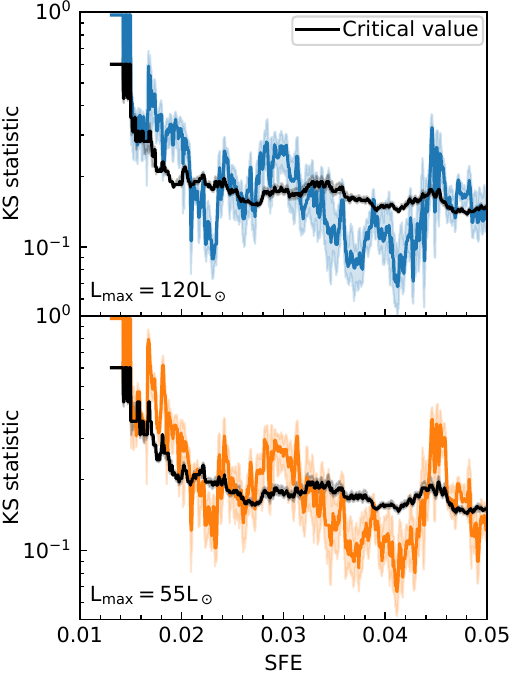}
    \caption{The smoothed KS statistic measured over SFE using the 2D-Bound settings (See \Cref{tab:comparison_summary}) for upper luminosity limits of $120\,\mathrm{L}_\odot$ (\emph{top}) and $55\,\mathrm{L}_\odot$ (\emph{bottom}). The smoothing window is SFE$=0.01\%$. The median value is plotted with an opaque line, and the shaded regions show the 16th and 84th percentile of the integrated values. The critical KS statistic (calculated with \Cref{eqn:KS_crit} is shown by the black line.}
    \label{fig:KS_test}
\end{figure}

We calculate the KS statistic between SFE=0.01 and 0.05, and we also calculate a corresponding critical value, below which, the two distributions are significantly similar. The critical value is found using:

\begin{equation}
    KS_{crit} = \sqrt{-ln\left(\frac{\alpha}{2}\right)\times\left(\frac{1+\frac{m}{n}}{2m}\right)}
    \label{eqn:KS_crit}
\end{equation}

\noindent where $m$ and $n$ are the number of values in each of the two samples (i.e. the number of separations, which is 39 for Perseus), and $\alpha$ is derived from the confidence level. We use $\alpha=0.99$, therefore if the KS statistic is below the calculated critical value, the two samples are consistent to a confidence level of 99\%. We calculate KS statistics for the number of visible stars using limits in 2D-Bound analysis, for both luminosity limits.

We smoothed the KS statistic and critical value over a window of SFE$=0.01\%$, and the results are shown in \Cref{fig:KS_test}. The shaded regions show the 16th and 84th percentile of the integrated values. For both luminosity limits, we see that bimodal distributions appear at various stages in the simulation. This is not surprising because we expect many systems to evolve from large to small separations, meaning that at some points in the simulation, the valley observed between $\sim300-1000\au$ is filled in.

Based on the KS test, the time when we produce a distribution that is most consistent with the Perseus observation is at approximately SFE$=4.1\%$ for both $L_{max}=55$ and $120\,\mathrm{L}_\odot$.

\begin{figure*}
    \includegraphics[width=\linewidth]{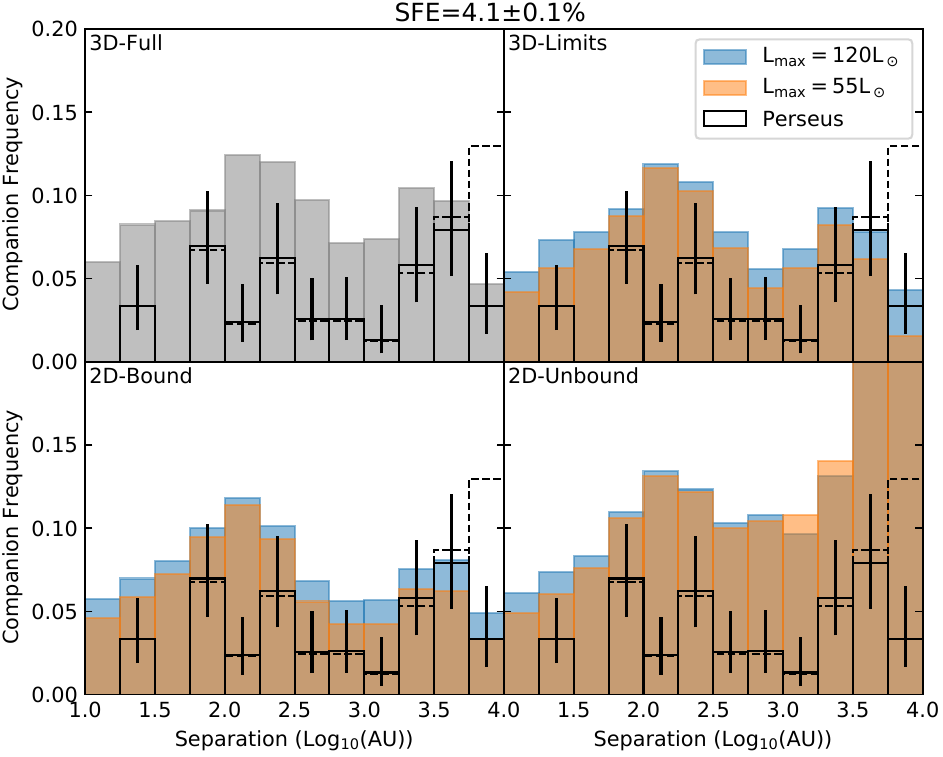}
    \caption{Same as \Cref{fig:CF_hist_42}, but for SFE=$4.1\%$}
    \label{fig:CF_hist_41}
\end{figure*}

\subsubsection{Best fit to observation}

Based on the time of best fit found by the two-sample KS tests, we reproduce \Cref{fig:CF_hist_42} for SFE=$4.1\%$. This is shown in \Cref{fig:CF_hist_41}. Although it is only slightly earlier than the CF versus separation histogram shown in \Cref{fig:CF_hist_42}, the bimodal distribution is more pronounced when luminosity limits are applied in 3D-Limits and 2D-Bound. While the values of the CF may vary between the simulations and observations, the positions of the peaks and the valley are in good agreement.

From this, we find that the observed bimodal distribution can be produced by multiple star formation via non-disc fragmentation pathways alone. This is not to say that disc fragmentation does not occur, but that a significant number of binaries with separations $20-100\au$ are probably formed on core fragmentation scales and migrate. The CF values from the simulation are generally higher than observations, and in order to reduce the overall CF value, more single stars are needed.

If our simulations resolved disc fragmentation we expected a higher CF at lower separations and this may shift the inner peak to lower separations.

\section{Discussion}
\label{sec:discussion}

\subsection{Star formation environment on formation pathways}

Our work specifically looked at whether multiple star formation on core fragmentation scales can reproduce observations of protostellar multiplicity. The minimum scale at which fragmentation occurs is set by the physics used, for example angular momentum, magnetic support, or tidal forces \citep{guszejnov_protostellar_2017, haugbolle_stellar_2018, lee_stellar_2019}. Our initial condition of an isothermal equation of state (EOS), turbulent driving, and ideal MHD produces a typical fragmentation scale of approximately $1000-3000\au$ (c.f. \Cref{fig:initial_separations} and \ref{fig:frag_scale}). \citet{guszejnov_effects_2023} ran cluster simulations with a Lagrangian code with similar initial conditions (turbulent initial velocity field, ideal MHD) but also employ radiation-hydrodynamics, found similar fragmentation scales of $1000-10000\au$. The fragmentation scale produced by our simulations and \citet{guszejnov_effects_2023} are consistent with observations of fragmentation in various star-forming regions having separations of a few thousand astronomical units \citep{palau_thermal_2018, figueira_ALMA_2018}.

In our work, we identified the unbound core fragmentation pathway, where a star is initially born unbound, and then later becomes bound to the star or system it was most bound to at birth. Other simulations have modelled the formation of systems that were initially unbound, but through accretion, and angular momentum exchange with the surrounding medium, they become bound \citep{ostriker_capture_1994, moeckel_capture-formed_2007, bate_stellar_2012, munoz_survival_2015}. This process has been called `gas-mediated capture', and this is likely to be occurring in our unbound core fragmentation cases. Previous simulations have concluded that this mechanism would not occur frequently. In our simulations, approximately 10 to 25\% of systems form via this mechanism (c.f. \Cref{fig:formation_pathway_frac}) with this pathway occurring more in low-density environments.

The dynamical capture pathway was thought to not occur often \citep{tohline_origin_2002}. However, simulations of clustered star formation and observations of young stars may suggest that flybys may occur frequently. Flybys by unbound stars have been proposed as the trigger of the excitation of spiral arms observed in some circumstellar discs \citet{perez_spiral_2016, cuello_dusty_2019}, and potentially trigger accretion bursts \citet{borchert_rise_2022}. \citet{pfalzner_close_2021} used N-body simulations of the stellar dynamics in young clusters and determined that stellar flybys are probably more common than initially expected in low mass clusters, and 10\%-15\% of discs should show evidence of this. \citet{pfalzner_close_2021} suggest that while the overall stellar density is lower in low mass star-forming regions, the distribution of stars is more clustered, which can aid stellar interactions in the central regions. In our work, we also confirm that star formation is more clustered in lower-density environments.

With the evidence that stellar flybys are not uncommon, it is not unreasonable to suggest that some flyby events may turn into dynamical capture events. Our work finds that for stars that are born unbound but later become bound, in higher-density environments, they are more likely to follow the dynamical capture pathway. \citet{murillo_siblings_2016} using SED fitting, inclination effects, and outflows, measure how coeval young multiple star systems in Perseus were. They found that approximately a third of the objects were non-coeval, suggesting these systems did not necessarily form together. This fraction is consistent with our $M_{gas}=3750\msun$ simulation which also found around a third of systems follows the dynamical capture formation pathway (see \Cref{fig:formation_pathway_frac}). Our simulations also suggested that for stars that are born unbound, the fraction that follow the dynamical capture pathways increases in higher-density environments.

Once a bound binary is formed, it can proceed to evolve to smaller separations via dynamical friction with the surrounding gas. This is because in the early stages when a star is low mass in a dense environment, the dynamics are dominated by the gas. The young star dynamics are influenced by the momentum gained via accretion, and the dense wakes that form from moving through dense gas mediums \citep{bate_accretion_1997, ostriker_dynamical_1999, stahler_orbital_2010, kuruwita_dependence_2020}. \citet{lee_formation_2019} ran simulations of clustered star formation using the MHD AMR \texttt{ORION2} with similar initial conditions to our simulations and derived a model to describe the orbital decay seen in two of the formed binaries. This model describes the angular momentum evolution of the simulated binaries using:

\begin{equation}
    \dot{L} = (\mathbf{r}_1 \times -\dot{m}_1 \mathbf{v}_{1,rel}) + (\mathbf{r}_2 \times -\dot{m}_2 \mathbf{v}_{2,rel})
    \label{eqn:lee_orb_evol}
\end{equation}

\noindent where $\mathbf{r}_*$ is the separation vector of the star to the centre of mass, and $\mathbf{v}_{*,rel}$ is the relative velocity between the star and the gas. The model describes the later evolution ($>100\kyr$) of the simulated binaries well and finds that orbital evolution is halted when the binary is no longer embedded because there is no gas around to create drag. This formulation suggested that with a higher accretion rate, the inspiral rate is faster.

In our work, we find that the characteristic initial inspiral rates ($<10 000\yr$, c.f. \Cref{fig:inspiral}) do not vary significantly between star-forming environments (i.e. different simulations). This inspiral behaviour may eventually reflect the evolution that is modelled by \citet{lee_formation_2019} at later times, but this is not measured in this work. Within each simulation, there is a substantial spread in the inspiral rates measured over our baselines. This may suggest that the orbital evolution is more sensitive to the local environment around the system than the overall star-forming environment. Curiously, we find that systems that formed via dynamical capture experienced faster inspiral rates when measured over the first $10\,000\yr$ of evolution since the first periastron. This was a surprise because we hypothesised that these systems would inspiral slowly. However, many of these systems may have experienced an interaction that quickly hardens previously unbound pairs.

It is difficult to observe evidence of orbital evolution in wide binaries ($a>1000\au$) due to the long orbital periods for individual systems. However, observations of OB associations find that the velocity dispersion increases with age \citep{ramirez-tannus_relation_2021}. This relationship between velocity dispersion and age is suggested to be caused by the hardening of binaries, supporting orbital migration forming close binaries. However, this velocity dispersion evolution is observed over millions of years, but simulations of binary star formation from core fragmentation frequently produce binaries that experience significant orbital evolution very early on as described above.

\subsection{Reproducing observations}

In \Cref{sec:obs_comparison} we investigated whether our simulations reproduced the bimodal CF versus separation distributions observed in protostars in Perseus \citep{tobin_vla_2016}. This bimodal distribution has also been observed in Orion \citep{tobin_vlaalma_2022}, and \citet{encalada_870_2021} appear to resolve the inner peak in Ophiuchus. The valley between the two peaks has also been observed in Class I objects in various star-forming regions across the sky by \citet{connelley_evolution_2008}. Because this bimodal distribution has been observed in multiple regions it is expected to be a long-lived feature.

One proposed hypothesis for the origin of a bimodal separation distribution comes from \citet{heggie_binary_1975} who suggested that hard binaries become harder and soft binaries become softer, leading to a bimodal separation distribution. A mechanism for orbital hardening is via the ejection of a companion. This companion can either be completely ejected from the system (becomes unbound) or pushed to a wider orbit. Many binaries in very wide orbits are found to be in triple systems, and it is hypothesised the outer companion was ejected to a wider orbit \citep{reipurth_formation_2012}. Ejections of companions and changing multiple star systems architectures also occur in gas-rich environments, like our own simulations and in other works \citep{bate_stellar_2012, ryu_analytic_2017}.

Conversely, when the protostellar bimodal distribution was first observed in the Perseus star-forming region, the two peaks were attributed to disc fragmentation producing close binaries, and core fragmentation producing wide binaries \citep{tobin_vla_2016}. However, as shown throughout this paper, binaries that are formed on core fragmentation scales often experience significant orbital evolution early in their formation. We calculated CF versus separation distribution of multiple star systems, looking at the true distribution, as well as applying observational limits. We successfully produce bimodal distributions at the calculated characteristic SFE of the observed objects in Perseus (4.2\%), and later find a better match to the \citet{tobin_vla_2016} observations at SFE = 4.1\%.

The origin of the bimodal distribution in our sims appears to be driven but the inspiral of young binaries, which migrate from separations of 100-1000s $\au$ to tens of AU. While there are complex N-body interactions occurring in our simulations, it does not seem that the mechanism described by \citet{heggie_binary_1975} is dominant, at least, not on the timescales explored in this work.

As stated previously, this observed bimodal distribution may be long-lived and it is not clear if the evolutionary state of a star-forming cloud would affect when this distribution is observed. Multiple star systems form throughout all of the simulations as shown in \Cref{fig:separation_vs_SFE}, therefore we expect to constantly observe systems forming at large separations ($>1000\au$) and in-spiralling to smaller separations. The derived best fit is close to the derived characteristic SFE, and it is not clear if this is coincidental.

While the best fit between simulations and observations occurs near when the SFE of the simulation matches observations, when we look at the calculated KS statistic against the critical value in \Cref{fig:KS_test}, we see that good statistical agreement appears throughout the evolution of the simulations. Looking at the evolution of the KS statistic, bimodal distributions with similar shapes to that observed in Perseus appear and disappear at different stages. There is an extended period where good statistical agreement is found between SFE $\sim$3.9-4.3\%, which spans $\sim$52$\,\mathrm{kyr}$ in simulation time. This bimodal distribution will be visible for approximately 10\% of the simulation run time. This makes it clear that the feature in our simulation is not spurious, but is a robust feature.

Overall, we can reproduce the observed bimodal distribution in protostellar CF using multiple star systems formed on core fragmentation scales only due to orbital migration. This is not to say that disc fragmentation does not occur, but that a significant fraction of closer binaries ($20-100\au$) can form via non-disc fragmentation pathways.

\section{Limitations and caveats}
\label{sec:caveats}

\subsection{Numerical resolution}

On the highest level of refinement, the resolution of our simulations is $50\au$, which means typical circumstellar disc sizes ($\sim$75$\au$; \citet{cox_protoplanetary_2017, ansdell_alma_2018}) are only resolved over a couple of cells. However, this work only investigates star formation from molecular cloud fragmentation, therefore, resolving discs is not crucial to this work.

Discs are an important part of the mechanism that determines how much mass is accreted onto stars, and what is ejected via outflows, and this is discussed in the next section.

The sink particle motion is calculated using a leap-frog integrator, and the smoothing length is a third of a cell length on the highest resolution, i.e. $16.6\au$. The sink particle motion is calculated using the gravitational potential from both other sink particles and the gas potential. The integrator can accurately calculate the N-body particles until the separation is near the softening length. However, the gravitational potential contribution from the gas can be less resolved as the two sink particles approach separations near the softening length.

Overall, the numerical resolution is limited near and sink particles, and may not accurately simulate orbital evolution when the separations are comparable to the softening length. The work in this paper has primarily focused on resolving the bimodal distribution observed in protostars, with the inner peak at projected separations of 75$\au$. With our simulation setup, we can sufficiently resolve this inner peak.

\subsection{Non-ideal MHD effects and resolving outflows}

Our simulations compute the ideal-magnetohydrodynamic equations. For molecular clouds, the typical fractional ionisation (abundance of electrons) is $\sim$10$^{-6}-10^{-8}$. With this ionisation fraction, ambipolar diffusion is present to help dissipate magnetic field flux to allow protostellar cores to collapse. While we do not explicitly include ambipolar diffusion in the computation of the MHD, \citet{hennebelle_role_2019} argue that the numerical diffusion naturally present in hydrodynamic simulation is sufficient to reproduce the effect of ambipolar diffusion. Other non-ideal effects are not dominant in the molecular cloud regime, however, \citep{wurster_there_2019} investigated full non-ideal MHD with clustered star formation with smoothed particle hydrodynamic simulations and found there is less magnetic breaking leading to discs forming more easily. With our resolution, we do not resolve discs, therefore, we do not expect the inclusion of other non-ideal effects to affect our results.

Magnetic fields are responsible for launching jets and outflows which return mass and momentum to the surrounding protostellar environment. With the resolution of $50\au$, jets are not self-consistently produced while weak outflows can be produced with larger discs. Not resolving outflows will not affect the motion of the sink particles significantly, but outflows regulate what fraction of mass is accreted onto the star and what is lost via outflows. As described in \Cref{ssec:sink_model}, we assume a mass accretion fraction of $50\%$. The fraction of mass that is accreted is not well constrained and various disc wind models give accretion fractions from 40-90\% \citep{seifried_magnetic_2012, fendt_bipolar_2013}. Observations of T-Tauri stars find mass accretion fractions of 50-99\% \citep{nisini_connection_2018}, and observations of Herbig-Haro objects find fractions of $\sim$90\% \citep{ellerbroek_outflow_2013}.

The accretion fraction used in our simulations may be on the lower end of typical accretion fractions but it is still consistent with models and observations.

\subsection{Radiation feedback}

An isothermal equation of state is used in our simulations, therefore, the internal energy remains constant throughout the simulations. This equation of state is satisfactory for our simulations because with our resolution we do not enter the regime where adiabatic heating of protostellar cores occurs ($\sim$3.8$\times$10$^{-13}\,\mathrm{g}\,\mathrm{cm}^{-3}$; \citet{masunaga_radiation_2000}). This is because the density threshold for sink particle formation does not exceed $1.3\times 10^{-14}\,\mathrm{g}\,\mathrm{cm}^{-3}$.

While an isothermal equation of state is justified in this work for the resolution at which we simulate hydrodynamics, the sink particles are used as proxies for stars, which would be able to produce radiation feedback. Radiation pressure from star formation can inject energy back into the star-forming environment, however, radiative MHD simulations carried out by \citet{rosen_role_2020} find that magnetic fields are dominant over radiation pressure feedback, even in massive stars. Observations also suggest that gas temperature may not have a strong impact on fragmentation and rather, mass and density are key factors in fragmentation \citet{murillo_role_2018}.

Radiation feedback from the protostars is also found to suppress disc fragmentation \citep{offner_effects_2011}. However, in this work, we are only concerned with multiple star formation on core fragmentation scales.

\section{Summary and conclusions}
\label{sec:conclusion}

We present a numerical study investigating protostellar multiplicity in star-forming environments of varying density. We find the following main results:

\emph{multiple star formation pathways}: Three main formation pathways were identified in this study, which were bound core fragmentation, unbound core fragmentation, and dynamical capture. Approximately a third of systems with a new star form via dynamical capture, independent of the star-forming environment. This is consistent with observations that find approximately a third of protostellar binaries appear to not be coeval \citep{murillo_siblings_2016}.

\emph{Orbital evolution}: We find that systems that form on core fragmentation scales undergo significant orbital evolution. The median inspiral rate when measured on $1\,000$ and $10\,000$ year baselines does not vary with star-forming environments. However, within a single simulation, the spread in inspiral rates is large, implying that the orbital evolution is probably most strongly affected by the local environment around the system, rather than the larger star-forming environment.

We find the inspiral rate when measured over $10\,000$ years does begin to segregate based on formation pathways, with systems that form via dynamical capture in-spiralling faster than those formed via bound core fragmentation. This suggests that systems that form via dynamical capture undergo an interaction that hardens the system quickly, and prolongs the inspiral process, while bound core fragmentation systems reach a steady orbit sooner.

\emph{Reproducing observed protostellar bimodal separation distribution}: We used the simulation with the global properties closest to the Perseus star-forming region, and derived companion frequency versus separation distributions. When selecting SFEs in the model that matches the observed SFE of Perseus, we find a similar bimodal distribution to the one observed in Perseus \citep{tobin_vla_2016, tobin_vlaalma_2022} when using the same selection criteria for the sink particles as in the observations. This is due to orbital migration from large separations to $20-100\au$. The valley between the peaks is less defined than in the observations and is not globally present throughout the evolution, but only in a range of SFEs from 3.9\% to 4.3\%.

Overall our results are compatible with core fragmentation and dynamical capture being the only sources of multiplicity for separations above $20\au$, without the need for any contribution from disc fragmentation. Multiple star formation on core fragmentation scales (100s to 1000s of AU) can follow different pathways, but many systems experience significant orbital evolution resulting in closer separations. The star formation environment may influence which pathways are more prominent, but the subsequent inspiral rate is independent of the global environment. The result of the orbital evolution is that multiple star systems with a range of separations are produced, including close binaries. This orbital evolution naturally leads to the observed bimodal distribution seen in the Perseus star-forming region when considering large enough SFEs.

\section*{Acknowledgements}

We thank the anonymous referee for their insightful comments and suggestions. The research leading to these results has received funding from the Independent Research Fund Denmark through grant No.~DFF 8021-00350B (TH, RLK). This project has received funding from the European Union’s Horizon 2020 research and innovation Program under the Marie Sklodowska-Curie grant agreement No. 847523 ‘INTERACTIONS’. The astrophysics HPC facility at the University of Copenhagen, supported by research grants from the Carlsberg, Novo, and Villum foundations, was used for carrying out the simulations and analysis, as well as long-term storage of the results. RLK also acknowledges funding from the Klaus Tschira Foundation. \texttt{yt} \citep{turk_yt:_2011} was used to help visualise and analyse these simulations.

\bibliographystyle{aa}
\bibliography{references.bib}

\clearpage
\onecolumn
\appendix
\section{Examples of formation pathways taken from simulation}
\label{sec:appendix}

\begin{figure*}[!b]
    \centerline{\includegraphics[width=1.0\linewidth]{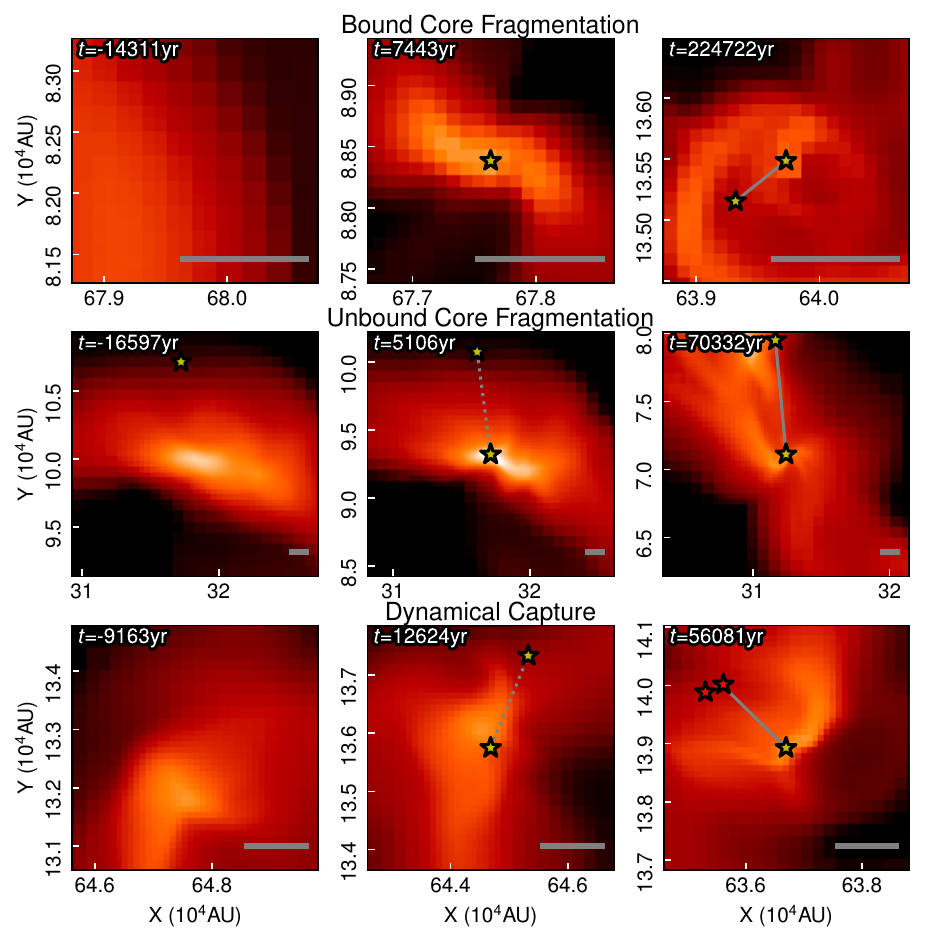}}
    \vspace{-0.5cm}
    \caption{Shows examples of the formation pathways described in \Cref{fig:formation_pathway_schematic} in \Cref{ssec:formation_pathways}. Projections were taken from candidates in the $M_{gas}=3000\msun$ simulation. The grey bar in the bottom right of each panel annotates $1000\au$. The time annotated in the top left of each panel indicates the time with respect to the star that is centred on for each pathway. }
    \label{fig:pathway_appendix}
    \vspace{-0.5cm}
\end{figure*}

\vspace{-0.5cm}
\begin{multicols}{2}

In \Cref{fig:pathway_appendix}, we show examples of the formation pathways described in \Cref{fig:formation_pathway_schematic} in \Cref{ssec:formation_pathways}. A system's formation pathway was determined by the sink particle data, which is written out at a $\sim$2$\yr$ cadence. Due to the coarse cadence of the hydrodynamic data ($\sim$22$\kyr$), not all systems could be plotted.

The projections are centred on the primary component in the bound core fragmentation pathway, and on the secondary companion in the other formation pathways. The position of the pre-sink formation frame (the projections in the left column) is estimated using the initial velocity of the centred star at birth. The projections are integrated over a box, where the side length is $2\times (r_{sep\_max}+500\au)$ in the xy-plane, where $r_{sep\_max}$ is the maximum separation between the centred sink particle and other sink particles related to the star.

The colour bar of each projection spans three orders of magnitude and is centred on the mean density in the projection. At the bottom of each panel, $1000\au$ is annotated by a grey bar. A solid line between stars indicates a bound relationship (i.e. $E_{pot} > E_{kin}$). A dashed line signifies a relationship where the system is unbound, but this is the system that the centred sink has the lowest energy relationship with.

The centred star and the sink particle or system that the centred star is most bound to at birth are annotated by yellow stars. Other stars are annotated in red.
\end{multicols}
\end{document}